\documentstyle[aps,twocolumn,psfig]{revtex}  

\newcommand{\th}{\vartheta}  
\newcommand{\f}{\varphi}  
  
\newcommand{\ga}{\gamma}  
\newcommand{\om}{\omega}

\begin{document}  
\author{V.I.Dimitrov\thanks{Permanent adress: Faculty of Physics, University of Sofia,  
1164 Sofia, Bulgaria}, S. Frauendorf and F. D\"onau}  
\address{Department of Physics, University of Notre Dame, Notre Dame, IN  
46556, USA \\  
and Institute for Nuclear and Hadronic Physics, Research Center Rossendorf, PB\\  
51 01 19, 01314 Dresden, Germany}  
\title{Chirality of nuclear rotation}  
\maketitle  
  
\begin{abstract}  
It is shown that the rotating mean field of triaxial nuclei can break the  
chiral symmetry. Two nearly degenerate   
$\Delta I =1 $ rotational bands originate from the  
left-handed and right-handed solution.  
\end{abstract}  
\vspace{0.1cm}  
\vspace{0.5cm}  
  
\begin{figure}[tbp]  
\psfig{file=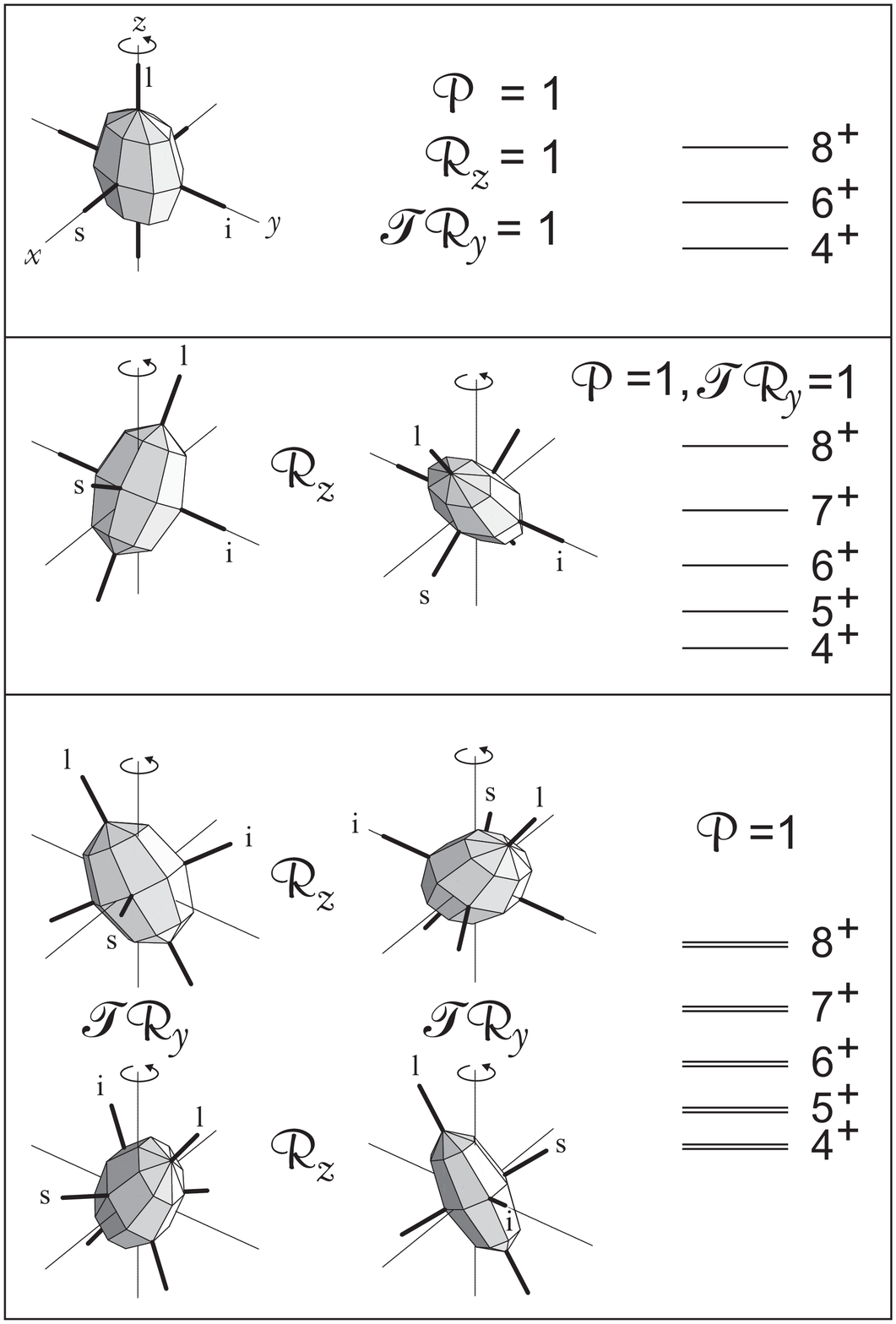,width=7cm}  
\caption{The discrete symmetries of the mean field of a rotating triaxial  
reflection symmetric nucleus (three mirror planes). The axis of rotation (z)  
is marked by the circular arrow. It coincides with the angular momentum\ $%
\vec J$. The structure of the rotational bands associated with each symmetry  
type is illustrated on the right side. Note the change of chirality induced  
by ${\cal T  R}_y(\pi)$ in the lowest panel.}  
\label{f:sym}  
\end{figure}

Chirality appears in molecules composed of more than four different atoms  
and is typical for the  biomolecules.   
In chemistry it is of static nature because it characterizes  
 the geometrical arrangement of the atoms.    
Particle physics is the other field where chirality is encountered.  
 Here it has a dynamical character, since  it distinguishes  
between the parallel and antiparallel orientation of the spin with respect  
to the momentum of massless fermions. 
Meng and Frauendorf \cite{chiral}  
recently pointed out that the rotation of triaxial nuclei may attain a   
chiral character. The lower panel of fig. \ref{f:sym} illustrates this new  
possibility. We denote the three principal   
axes (PA) of triaxial density  
distribution   
by l, i, and s, which stand for long, intermediate and short,  
respectively. The angular momentum\ vector $\vec J$ introduces chirality by  
selecting one of the octants. In four of the octants  
 the axes l, i, and s form a  
left-handed and in the other four a right-handed system. This gives rise to  
two degenerate rotational bands because all octants are   
energetically equivalent. Hence  
 the  chirality of nuclear rotation   
results from a combination of dynamics (the  
angular momentum) and geometry (the triaxial shape).  
  
 Our symmetry argument is based on the  
presumption that the axis of uniform rotation needs not to agree with one of  
the PA of the density distribution. This does not hold for a rigid triaxial  
body, like  a molecule for example, which can uniformly rotate  only   
about the l- and s-axes. However, already Rieman \cite{Riemann} pointed  
out that a liquid may uniformly rotate about an axis tilted with  
respect to the PA, if there is an intrinsic vortical motion. In the  
case of the nucleus the quantization of the angular momentum\ of the  
nucleons at the Fermi surface generates the vorticity which enables rotation  
about a tilted axis.  
  
The semiclassical mean field description of tilted nuclear rotation   
was developed  
in \cite{Kerman,Frisk,tac}. In the following we shall refer to it as  
the Tilted Axis Cranking (TAC) approach \cite{tac}. 
Fig. \ref{f:sym} illustrates the different  
symmetries  if  the mean field is assumed to be reflection symmetric. In the  
upper panel the axis of rotation (which is chosen to be z) coincides with  
one of the PA, i. e. the finite rotation ${\cal R}_z(\pi)=1$. This symmetry  
implies the signature quantum number $\alpha$, which restricts the total  
angular momentum\ to the values $I=\alpha+2n$, with $n$ integer  
($\Delta I=2$ band) \cite{bf}.  
In the middle panel the rotational axis lies in one of the planes spanned by  
two PA (planar tilt). Since   
then ${\cal R}_z(\pi)\not=1$, there is no longer a  
restriction of the values $I$ can take. The band is a sequence of states the  
$I$ of which differ by 1 ($\Delta I =1$ band). There is a second symmetry in  
the upper two panels: The rotation 
${\cal R}_y(\pi)$ transforms the  
density into an identical position but changes the sign of the angular  
momentum\ vector $\vec J$. Since the latter is odd under the time reversal  
operation ${\cal T}$, the combination ${\cal T R}_y(\pi)=1$.  
  
In the lower panel the axis of rotation is out of the three planes spanned  
by the PA. The operation ${\cal T R}_y(\pi)\not=1$. It changes the  
chirality of the axes l, i and s with respect to the axis of rotation $\vec J  
$. Since  the left- and the right-handed 
solutions  have a the same  
energy, they  give rise to two degenerate $\Delta I =1$ bands. They are the  
even ($|+\rangle$)  and odd ($|-\rangle$) linear combinations of the two 
chiralites, which restore  the spontaneously broken  
${\cal T R}_y(\pi)$ symmetry.   
 
Ref. \cite{chiral} studied  
 a proton  
particle and a neutron hole coupled to a triaxial rotor with maximal  
asymmetry ($\gamma =30^{o}$) and the irrotational flow  
relation ${\cal J}%
_{l}={\cal J}_{s}<{\cal J}_{i}$ between the moments of inertia. 
The  solutions of this model case describe the dynamics of the  
orientation of $\vec J$. It was found 
that at  the   beginning 
 of the bands the angular momentum originates from the particle and the hole,  
whose individual angular momenta are aligned with the s- and l-axes. These  
orientations correspond to a maximal overlap of the particle and hole  
densities with the triaxial potential. A $\Delta I=1$ band is generated by  
adding the rotor momentum $\vec{R}$ in the s-l plane (planar tilt). There is  
a second $\Delta I=1$ band representing a vibration of $\vec{J}$ out of the  
s-l plane. Higher in the band, $\vec{R}$ reorients towards the i-axis, which  
has the maximal moment of inertia. The left- and the right-handed positions  
of $\vec{J}$ separate. Since they couple only by some tunneling,  
the two bands come very close together  
in energy.  
  
The model study assumed: i) a triaxial shape with $\gamma \approx 30^{o}$,  
ii) a combination of excited  
high-j protons and high-j neutron holes   
and iii) ${\cal J}_{l}={\cal J}_{s}<{\cal J}_{i}$.  
In order to decide 
 if i)-iii) are simultaneously realized in actual  
nuclei we developed a 3D cranking code,  
which permits a microscopic study. In essence, it is the mean field   
approximation to  
the two body Routhian $H^{\prime }=t+v_{12}-\omega \hat{J}_{z}$, where any  
orientation of the deformed average  potential $V_{def}$ is permitted   
(cf. \cite  
{Kerman,Frisk,tac}). Only reflection symmetry is demanded. In the actual  
calculation the standard shell correction approximation is used (cf. \cite  
{NilsRag}). The quasi particle Routhian  
\begin{eqnarray}  
h^{\prime }=h_{sph}+V_{def}(\varepsilon ,\gamma )-\Delta \left(  
P+P^{+}\right) -\lambda N &&  \nonumber \\  
-\omega \left( \hat{J}_{1}\sin \vartheta \sin \varphi +\hat{J}_{2}\sin  
\vartheta \cos \varphi +\hat{J}_{3}\cos \vartheta \right) &&  
\end{eqnarray}  
is diagonalized. The monopole pair field $\Delta \left( P+P^{+}\right) $ and the term  
$\lambda N$, which controls the particle number $N$,   
are understood as sums of the neutron and proton parts. This  
yields quasi-particle energies and states, from which the the quasi  
particle\ configuration $|\omega ,\varepsilon ,\gamma ,\vartheta ,\varphi >$  
is constructed. The spherical Woods-Saxon energies \cite{Ref13} are used for  
$h_{sph}$ and the Nilsson model \cite{NilsRag} for $V_{def}(\varepsilon  
,\gamma )$. In a separate paper \cite{ba128} we demonstrate that this hybrid  
approximates well the deformed Woods Saxon potential, which provides a good  
description of the rotational spectra of the region around mass $A=135$.  
  
The total Routhian is calculated by means of the shell correction method,  
\begin{eqnarray}  
E^{\prime }(\omega ,\varepsilon ,\gamma ,\vartheta ,\varphi )=<\omega  
,\varepsilon ,\gamma ,\vartheta ,\varphi |h^{\prime }|\omega ,\varepsilon  
,\gamma ,\vartheta > &&  \nonumber \\  
-<\omega =0,\varepsilon ,\gamma |h^{\prime }|\omega =0,\varepsilon ,\gamma >  
&&  \nonumber \\  
+E_{LD}(\varepsilon ,\gamma )-E_{smooth}(\varepsilon ,\gamma ), &&  
\end{eqnarray}  
where $E_{smooth}$ and $E_{LD}$ are, respectively, the Strutinsky averaged  
and liquid drop energies at $\omega =0$. The method is described for example  
in \cite{NilsRag}. The deformation parameters $\varepsilon $ and $\gamma $  
and the tilt angles $\vartheta $ and $\varphi $ are found for a given  
frequency $\omega $ and fixed configuration\ by minimizing the total  
Routhian $E^{\prime }$. The value of $\Delta $ is kept fixed at 80\% of the  
experimental odd-even mass difference for the nucleus of interest.  
The values of $\lambda_p$ and and $\lambda_n$ are adjusted to have  
the correct values of   
$<Z>$ and $<N>$ for $\om=0$ and kept fixed for $\om>0$.

The electro-magnetic transition matrix elements  
for the left- and right-handed solutions  
 are obtained by means of  
the straightforward generalization of the semiclassical expressions  
given in \cite{chiral,tac,zphys}.  
If  tunneling is negligible 
there are no matrixelements between the two  
TAC states $|\om>$ and ${\cal T R}_y(\pi)|\om>$, 
because  
the electro-magnetic field cannot provide  
the amount of angular momentum which is necessary to reorient the long   
vector $\vec J$. 
Then the probabilities for the transitions  
$+\rightarrow +$ and $-\rightarrow -$ are given by the mean value 
and  the ones for $+\rightarrow -$ and $-\rightarrow +$ by half the 
difference of the transition matrix elements of  
the left- and the right-handed solutions. We assume zero tunneling. 
 
  
\begin{center}  
\begin{figure}[tbp]  
\psfull  
\vspace*{-2cm}  
\psfig{file=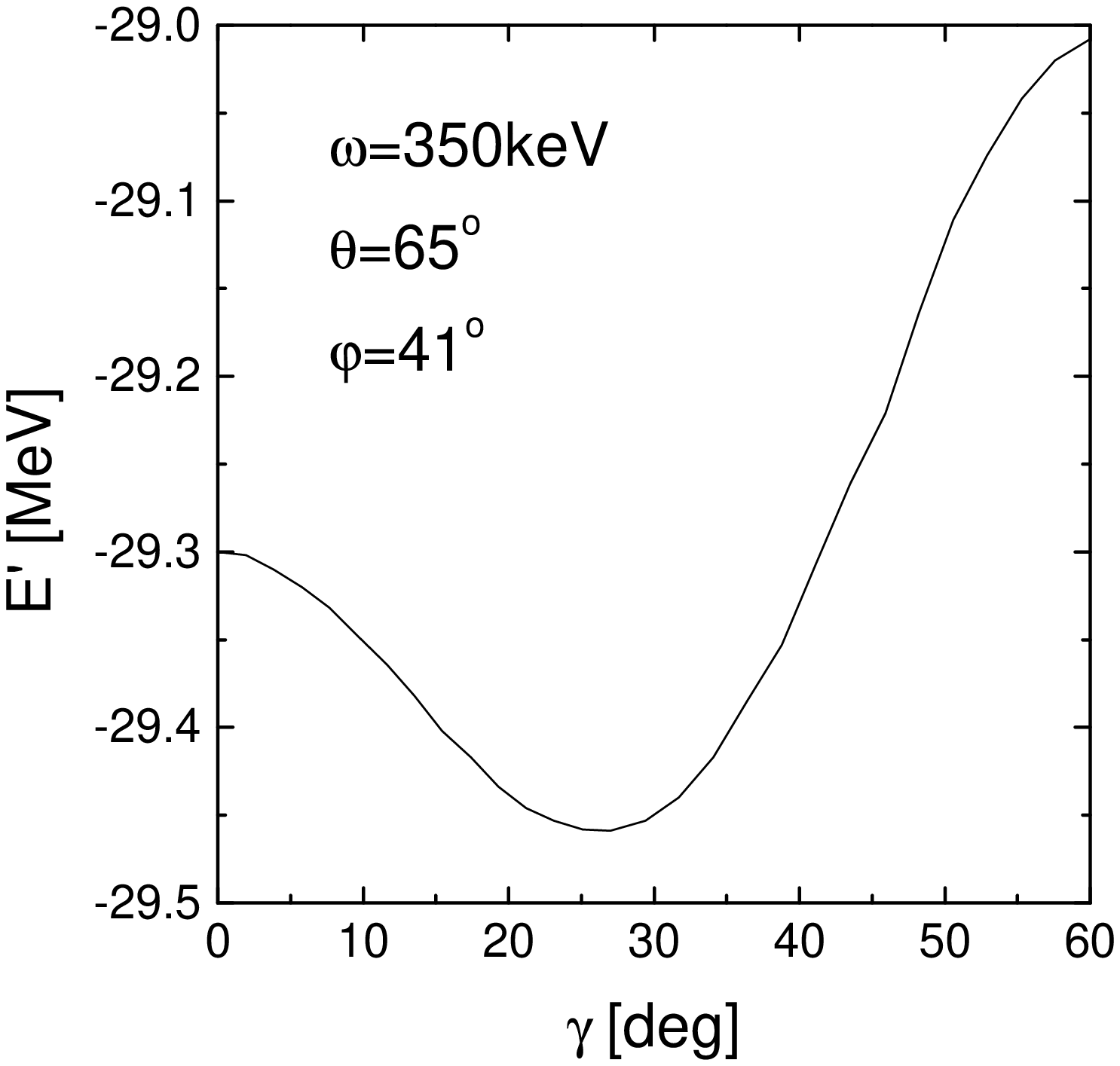,width=5cm}  
\caption{Total Routhian $E^{\prime }(\omega =0.35MeV,\varepsilon  
=0.175,\th =65^{o},\varphi =41^{o})$ as function of the triaxiality  
parameter $\gamma .$}  
\label{f:g350}  
\end{figure}  
\end{center}  
  
\begin{figure}[tbp]  
\psfull  
\psfig{file=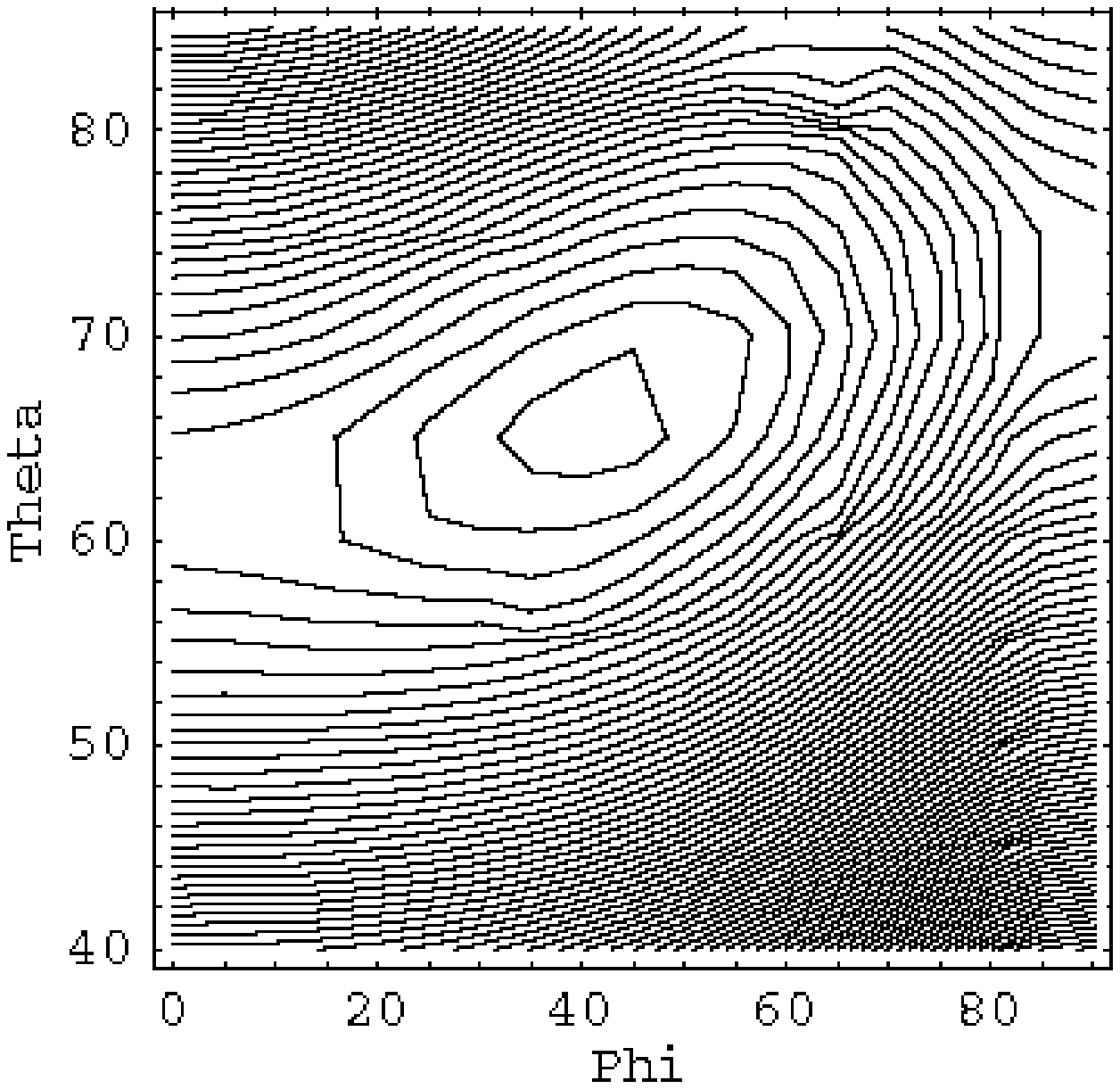,width=7cm}  
\caption{Total Routhian $E^{\prime }(\omega =0.35MeV,\varepsilon  
=0.175,\gamma =27^{o})$ as function of tilt angles $\theta $ and $\varphi $.  
The energy difference between the contour lines is 18 $keV.$ A cylindric  
projection is used. The PA correspond to $l:$ $\th =90^{o};$ $%
i:\th =90^{o},\varphi =90^{o};$ $s:\th =0^{o},\varphi =0^{o}$}  
\label{f:surf}  
\end{figure}

\begin{center}  
\begin{figure}  
\psfull  
\vspace*{-1.5cm}  
\psfig{file=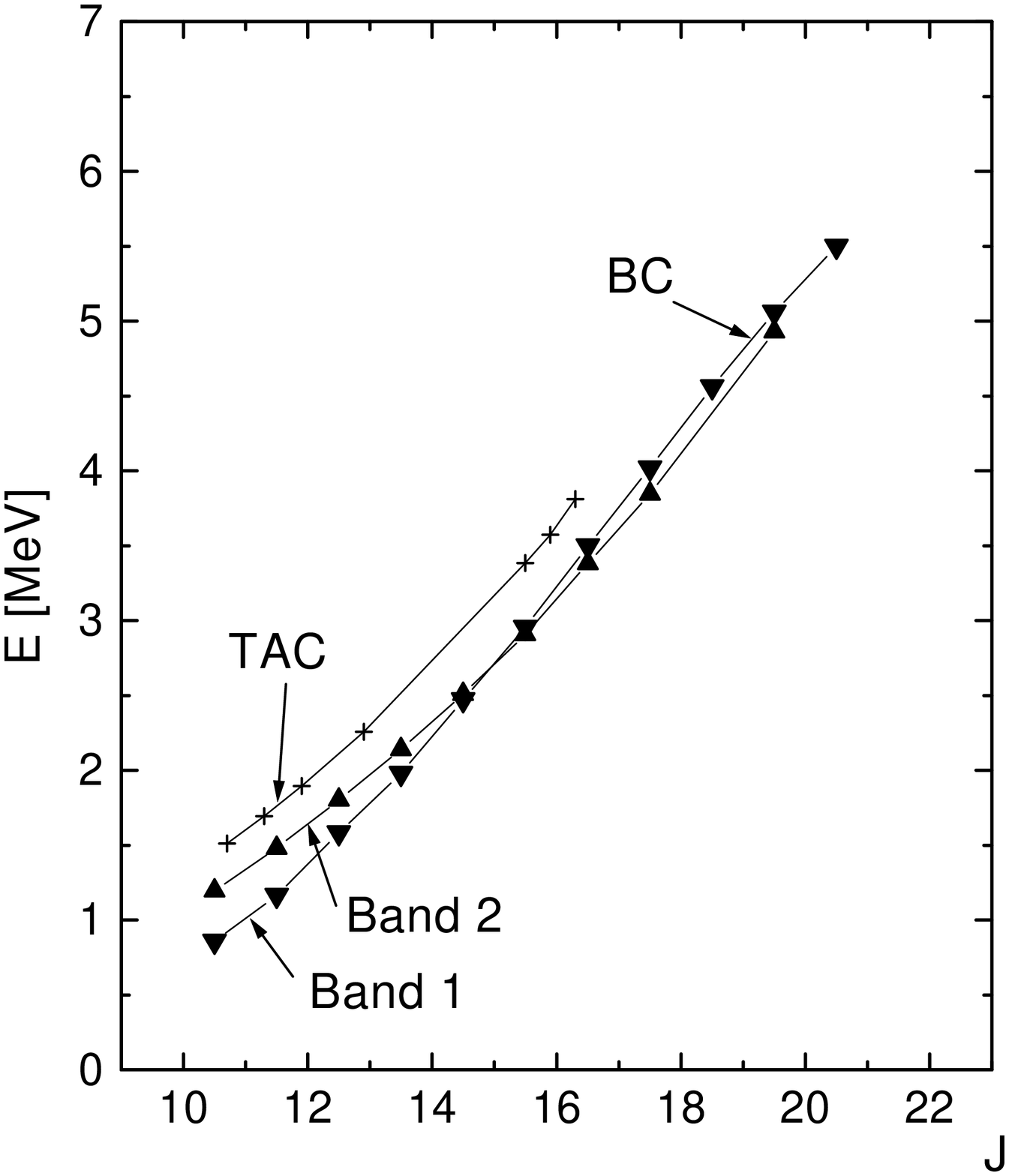,width=7cm}  
\vspace*{-0.5cm}  
\caption{Energy of the chiral bands   
in $_{59}^{134}$Pr$_{75}.$ Triangles: Experimental energies  
\protect\cite{pr134}  of the  
two bands assigned to the   
 configuration $[\pi h_{11/2},\nu h_{11/2}]$ vs. $J=I+1/2$.   
Crosses: TAC calculation (the zero point is arbitrary)    
.}  
\label{f:ei}  
\end{figure}  
\end{center}

\vspace*{-.8cm}  
We investigated the quasi particle configuration 
$[\pi h_{11/2},\nu h_{11/2}]$  
in $_{59}^{134}$Pr$_{75}$, which was suggested as a candidate for chiral  
rotation \cite{chiral}. The microscopic calculations,   
presented in figs. \ref{f:g350} - \ref{f:ei} as well as in the table,  
confirm the  mechanism for the chirality as described  
in the model study \cite{chiral}:

 - The shape is   
triaxially deformed with $\gamma$ close to $30^o$ (cf. Fig. \ref{f:g350}).    
The deformation remains nearly the same within the  
frequency range $0.3~MeV\leq \omega \leq 0.5$ $MeV$ studied.  
  
-The $h_{11/2}$ quasi proton has particle character and tends to  
align   with the s-axis. The $h_{11/2}$ quasi neutron has hole character and   
prefers the l-axis.     
  
-The moment of inertia ${\cal J}_i$ is the largest. Accordingly,   
the orientation of $\vec J$ ( position of the minimum in the $\th - \f$ plane)  
moves from the s-l plane ($\th=60^o,~\f=0^o$) towards the i-axis   
($\th=90^o,~\f=90^o$). Fig. \ref{f:surf} shows the energy  
in $\th - \f$ plane  at   
$\om=0.35~MeV$ when the orientation is furthest out   
in the chiral sector.  The minimum  is  soft in $\varphi $ direction.

\vspace*{-0.5cm}  
The  TAC mean field  approach  treats the  
orientation of the rotational axis in a static way.  
One can only give a simple interpretation of the  
two distinct cases of well conserved  ${\cal T R}_y(\pi)$-symmetry   
 (two separated $\Delta I=1$ bands) and strongly broken   
${\cal T R}_y(\pi)$-symmetry (two   
degenerate $\Delta I=1$ bands).  
For $\omega < 0.3~MeV$,  
 $\f=0$, i. e.   
 the rotational axis lies in the s-l plane.  
 For $\omega = 0.3~MeV$, the barrier at $\f= 0$ is only a few $keV$.   
Accordingly, two separate $\Delta I=1$ bands are expected, which 
  correspond to the first two vibrational  
states of the angular momentum vector $\vec{J}$.  
For $\omega =0.325 - 0.375~MeV$, 
the rotation is chiral and the two bands come close.  
Strong degeneracy is not expected, because the barrier at $\f= 0$ is only  
$60~keV$.      
For larger $\omega$, the minimum moves slowly   
towards the i-axis ($\th=90^o,~\f=90^o$). 
Fig. \ref{f:ei} shows the experimental energies of two lowest $\Delta I=1$  
bands of positive parity, which were suggested to form a chiral pair \cite  
{chiral}.  Their relative position correlates with the calculation:  
 At $I=10$, there is a substantial  
splitting between them,   
which becomes very small around $I=15$, and remains small for higher $I$.

\vspace*{-1cm}  
   
For given $\om$, band 1 has about two units of  angular momentum more than band  
2 (cf. fig. 6b in \cite{pr134}). The difference indicates 
substantial tunneling, which 
is expected because the calculated   
energy barrier between the two solutions    
is less than 100 $keV$ in $\f$ direction and the depth of the energy minimum   
with respect to the triaxiality parameter $\ga$ is only 180 $keV$. The  
situation is similar to the weak breaking of the reflection symmetry in   
the light actinide nuclei (see \cite{octupole}). There, the  barrier between the two   
reflection symmetric solutions is calculated to be few 100 $keV$. The two  
bands of opposite parity, which should be   
 
degenerate for strongly broken  
reflection symmetry, come fairly close in energy. But usually there is a  
difference between the angular momenta at given frequency  
  
 which is of the   
order of 1 - 2 units in most of the observed frequency range (1.3 for  
$^{226}$Th). Since the TAC mean field calculations do not include tunneling 
they must be compared with  
 the averages over the two chiral sister-bands.   
 The table demonstrates that the  
calculated curve $J(\omega )$ reproduces quite well  the average $\bar{J}%
(\omega )$ of the two experimental curves.    
 
Around $J=14$, the calculated ratio of 
$B(M1)/B(E2)\approx 3 (\mu_N)^2/(eb)^2]$  
for intraband transitions is consistent with  
the mean  of the experimental values of  
 $5\pm 3(\mu_N)^2/(eb)^2]$ \cite{pr134}. 
The experimental branching ratios  $B(M1,I_2\rightarrow I_1-1)/  
B(M1,I_2\rightarrow I_2-1)= 0.1,~0.1,~0.8,~0.5$   for $I=  
13,~14,~14,~16$ \cite{pr134} correlate with the calculated  increase of the  interband M1 transition 
probability when $\f$ becomes large.  
The experimental ratios for lower $I$ cannot be compared with the TAC calculations because the 
underlying assumption of no tunneling between the left- and right-handed 
solutions does not hold for small $\f$.  
The data does not permit us checking 
the characteristic small ratios $B(E2,I_2\rightarrow I_1-2)/  
B(E2,I_2\rightarrow I_2-2)$.

In the calculation, the one quasi neutron configuration [$h_{11/2}$]    
is crossed by the three quasi neutron configuration   
[$h_{11/2}^3$]  
near $\om=0.4~MeV$. We ignore this crossing   
 ("BC-crossing" in the common letter convention \cite{bf}), showing   
the results for the  [$h_{11/2}$] configuration.  
The reason is that in experiment the  crossing is   
observed at the higher  frequency of $\om=0.5~MeV$ \cite{pr134}.  
This kind of delay appears systematically in configurations with protons and  
neutrons in the same j-shell and is not accounted for by the Cranking model  
(see  \cite{pnodd} and the literature cited therein).

As a second example for chirality,  
we found the configuration\ $[\pi i_{13/2},\nu i_{13/2}]$ in $^{188}_{77}$Ir$%
_{111}$  (cf. table). The   
equilibrium shape  
is $\varepsilon=0.21$ and $\gamma=40^{o}$.  
Like for   $^{134}$Pr  we encountered the  BC-crossing, 
 which leads to the planar solution at $\om=0.3~MeV$. 
The existence of a chiral solution based on the excited  
  configuration [$\nu i_{13/2}$],  analogous  
to the one based on [$\nu h_{11/2}$] in $^{134}$Pr,   
is likely but we did not calculate it.  
 Ref. \cite{Frisk} found a chiral angular  
momentum\ geometry for the  configuration\ $[\pi g_{9/2},\nu g_{9/2}]$ in $%
^{84}_{39}$Y$_{45}$ {\em assuming} a triaxial shape with $\gamma =30^o$.  
We calculated for  this configuration  
an axial equilibrium shape   
with $\varepsilon = 0.19, \ga=60^o$ and  $\vartheta = 90^o$. This result is  
consistent with the observation of an even-$I$ yrast band (case of good  
${\cal R}_z(\pi)$-symmetry, upper panel of fig. \ref{f:sym}).  
  
In conclusion: We demonstrated the existence of self-consistent rotating  
mean field solutions with chiral character for   
$^{134}$Pr and $^{188}$Ir. The left- and right-handed  
solutions give rise to two near degenerate $\Delta I=1$ bands of the same  
parity with supressed electric quadrupole  transitions between them.   
A pair of bands with the expected properties is  observed in $^{134}$Pr.  
The calculated small energy gain due to breaking of the chiral symmetry  
 is in in accordance with the experiment, which points to substantial  
tunneling beween the two solutions of opposite chirality.   
  
This work was supported by the DOE grant DE-FG02-95ER40934.

\begin{table}[tbp]  
\begin{tabular}{lllllll}  
$\omega[MeV]$ & $\vartheta$ & $\varphi$ & $J$ & $\bar J$ & $B(M1)[(\mu_N)^2]$  
& $B(E2)[0.1(eb)^2]$ \\ \hline  
0.250 & 60 &  0 & 10.2 &      & 1.29~~~0.00 & 0.67~~~0.00\\ 
0.300 & 60 & 21 & 10.7 & 10.9 & 1.16~~~0.13 & 0.91~~~0.08\\ 
0.325 & 62 & 33 & 11.3 & 11.5 & 0.96~~~0.30 & 1.50~~~0.13\\ 
0.350 & 65 & 41 & 11.9 & 12.2 & 0.80~~~0.41 & 2.11~~~0.13\\ 
0.375 & 70 & 54 & 12.9 & 13.0 & 0.54~~~0.52 & 3.02~~~0.08\\ 
0.400 & 75 & 66 & 14.4 & 13.6 & 0.35~~~0.57 & 3.96~~~0.03\\ 
0.450 & 75 & 65 & 15.5 & 14.8 & 0.37~~~0.41 & 3.90~~~0.03\\ 
0.475 & 75 & 68 & 15.9 & 15.3 & 0.37~~~0.34 & 4.06~~~0.02\\ 
\hline  
0.200 & 66 & 13 & 12.2 &  & 1.43~~~0.23 & 1.35~~~0.16 \\  
0.225 & 66 & 25 & 12.7 &  & 1.25~~~0.83 & 2.29~~~0.53 \\ 
0.250 & 66 & 32 & 13.4 &  & 1.12~~~1.23 & 3.27~~~0.74 \\  
0.275 & 72 & 44 & 14.7 &  & 0.60~~~2.13 & 6.55~~~0.53 \\  
0.300 & 85 & 59 & 20.2 &  & 0.04~~~3.49 & 1.12~~~0.03 \\  
  
\end{tabular}  
\caption{Orientation angles $\vartheta$, $\varphi$ and value $J$ of the  
angular momentum and reduced transition probabilites of the configuration\ $%
[\pi h_{11/2},\nu h_{11/2}]$ in $^{134}_{59}$Pr$_{75}$ (upper part) and of  
the configuration\ $[\pi i_{13/2},\nu i_{13/2}]$ in $^{188}_{77}$Ir$_{111}$  
(lower part) as calculated by TAC.  
Chirality corresponds to   
both $\vartheta\not=0^o$ and $90^o$ and $\varphi\not=0^o$  
and $90^o$.  The experimental values $\bar  
J(\om)=(I_1(\om)+I_2(\om))/2+1/2$, where $I_{1,2}(\om)$ are obtained by  
interpolating between the discrete points for band 1 and 2, respectively.  
For the transition probabilities the left column 
gives the intraband transitions  
($+\rightarrow +$ and $-\rightarrow -$) and the right the interband 
transitions ($+\rightarrow -$ and $-\rightarrow +$). 
When the chiral symmetry is conserved, the intraband probabiites  are 
 equal to the non zero value. 
} 
\label{tab}  
\end{table}  
  
\end{document}